# On Simulating Processor Schedules and Network Protocols within CPS using TrueTime

Sreram Balasubramaniyan
Department of Information Technology
Kalasalingam University
Srivilliputhur, India
bsreram85@gmail.com

R.C.Hemesh
Department of Electrical and Electronics Engineering
Kalasalingam University
Srivilliputhur, India
kirthihemesh@gmail.com

B.Subathra
Department of ICE
Kalasalingam University
Srivilliputhur, India
clk0602@gmail.com

Saravanakumar Gurusamy
Department of ICE
Kalasalingam University
Srivilliputhur, India
saravana.control@gmail.com

Seshadri Srinivasan
International Research Center
Kalasalingam University
Srivilliputhur, India
seshucontrol@gmail.com

*Abstract*—Studying cyber-physical system (CPS) for a given network protocol and processor schedules is a challenging task. This investigation illustrates the role of TrueTime a MATLAB package for simulating CPS encapsulating information on processor schedules, and network protocols. Properties of CPS such as temporal behaviors, performance and stability can be studied using the TrueTime tool. Furthermore, these simulations can be used to benchmark design and propose design alternatives. The procedure to use the toolbox for simulations and instantiate various blocks to study CPS is illustrated in the paper. Finally, an illustrative example has been presented of CPS simulation with wireless network. Our example demonstrates that TrueTime is a useful tool for studying CPS performance, timing behavior, schedulability analysis, and studying suitable network protocols.

*Keywords*— Cyber-Physical System (CPS), network protocol, Processor Scheduling, TrueTime, simulations, Timing behavior, CPS performance

## I. INTRODUCTION

Cyber-Physical Systems (CPS) deals with integration of cyber world with the physical systems [12,33]. They are touted to be future of engineering systems and have found applications in wider domains such as automated highway systems [1], industrial automation [3], Intelligent Transportation [2], and smart energy systems to name a few. The performance of the CPS is affected by the operating principles of the cyber components such as scheduling and network protocols. Therefore, to design and benchmark CPS, its performance has to be tested with these components during design stage. As physical testing is limited due to cost considerations, simulation have emerged as promising solutions. Simulation tools that can be used for studying timing behavior, stability, and performance are required for designing CPS and performing design modifications. Furthermore, such simulation tools can be used for bench-marking CPS designs. TrueTime proposed by A. Cervin [24] is a tool that can be used for studying performance of control systems with a given network protocol and processor scheduling. This investigation studies the role of TrueTime in simulating CPS. In particular, simulation of CPS with timing discrepancy for a given network protocol, and process scheduling is studied. Furthermore, validation of controller design for a given performance specifications using TrueTime is also studied.

In literature, several tools and methodologies are used for studying the CPS performance that include specific aspect of CPS such as network performance, processor scheduling, or physical system properties. For instance, Network Simulator NS-2 [28], has the capability to add different protocols for all combinations of networks [34]. Visim [26] is another tool that uses NS-2 internally, but simplifies the task of simulating Mobile ad-hoc Wireless Network routing protocols. Authors demonstrated the performance of different routing prtocols. It is much easier to use and adapt than NS-2.

Formal models based on timed-automata for modeling the timing interfaces and then verified the timing performances using UPPAAL [31] have been studied in [3, 19]. Sreram [9] presented a simultaneous co-design and verification approach which utilizes the UPPAAL tool to verify the timed automata of a system having heterogeneous components. TimesTool [30] is another tool to analyze the dynamic behavior of the system and verify the task parameters and a given scheduling policy. It can also verify all the instances of the future states if the given system is schedulable.

Ptolemy [27] studies the functional model, and helps in analysis of real-time embedded systems and concentrates mainly on concurrent components, while metroII describes the architectural model [7]. Besides facilitating hierarchical models, the segregation in the architectural control and data flow between individual components enables the reliability and more usage in future models. Dymola [29] is a tool available open source for the modelling and analysis of integrated and complex systems, having the capability to simulate complex systems with multiple Functional Mockup Units. However, these tools can only study specific aspect of CPS such as network scheduling, timing analysis, verification or physical simulations. Designing CPS requires simulating the physical system considering the operating principles of the cyber components such as network protocol and processor schedules. TrueTime proposed by A. Cervin et al. [8] can be used to study the performance of CPS alongside the operating principles of cyber components. The role of TrueTime for designing CPS has not been completely exploited in literature.

TrueTime is a simulation tool for controller task execution in real-time kernels, network transmissions, and continuous plant dynamics [35]. TrueTime simulates the real time behavior of multi-tasking real-time kernels containing controller tasks. It can also be used to control behaviour on CPU and network scheduling . The support of Battery enables TrueTime to support physical systems which is operated on a separate power source [4, 5]. The current version 2.0 in beta release, introduces several new features including ultrasound network and modifying the kernel block to introduce automatic connections using hidden goto block so that the receive and send block can be called internally from the configurations that are made in block dialogues. This removes the necessity of connections between the kernel and the network blocks that were mandatory earlier. Moreover, for wireless networks this enables easy transfer of data packets and mimics the real world scenario.

Main contribution in this work is to analyse the role of TrueTime in simulating CPS with timing imperfections in network and processors [36]. The network has imperfections such as network delays [14], temporal delays in processor schedules etc. In [3], it has been shown that most of these timing uncertainties can be modelled as delays. These delays have been modelled using empirical distribution [16, 17], Markov Chain [15], Markov Chain Monte Carlo [13], non-liner models [20]. This investigation extends the delay models within CPS and then tries to simulate CPS performance with network imperfections. The method to simulate and study the behavior are also presented.

The investigation is divided into following sections. Section II deals with important blocks and its functions and parameters. Section III shows the implementation of TrueTime with an example and describes it with appropriate results and how to utilize the same effectively for validation of performance and stability of systems with timing imperfections and variable delay. Section IV concludes the paper by discussing the salient features of TrueTime.

II. SIMULATION OF CPS IN TRUETIME

Every TrueTime toolbox simulation scheme should contain three crucial parts: TrueTime kernel (computer, I/O device or some embedded system), TrueTime network (network model) and a controlled process. There is also an optional part TrueTime battery. TrueTime kernel is responsible for I/O and network data acquisition or data processing and calculations. It can realize a control algorithm/logic and it is the "brain" of every device. It uses several simple M-files (modified by us to satisfy our needs) which handle all mentioned operations. In the kernel can be executed several independent tasks (periodic, non-periodic) which can cooperate on the same goal. If we include TrueTime battery block we can set power supply for true time kernels.

A. Detailed Description of TrueTime blocks

Several blocks contribute towards the simulation of Cyber Physical system. TrueTime Library is shown in Fig. 1, and important blocks being detailed as below.

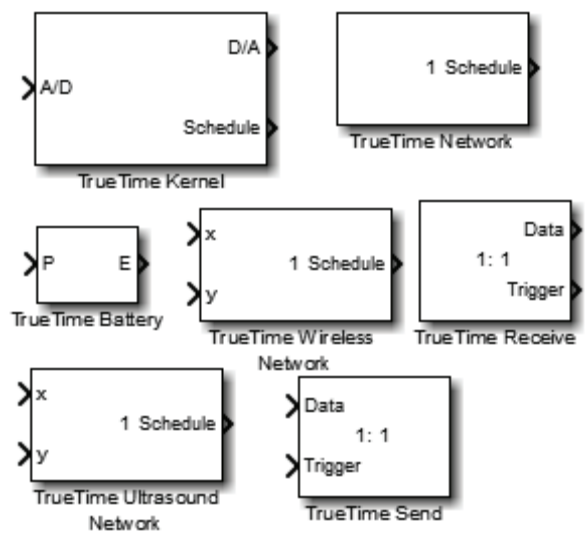

Fig. 1. TrueTime Library

*1) Kernel block:* The core block of TrueTime helps to set the initialization file which can be used not only to specify and set the scheduler parameters but also the interrupt handlers, and power management block handlers. These can be configured as separate files which helps in modularity and custom configurations of individual handlers. The Kernel block can be used to simulate sensors, actuators and controllers. One additional advantage of the kernel block is that the functionality of the Sensor/Actuator/Controller blocks can be simulated in a single block or can be simulated in combination as shown in Section III. Kernel block has the following parameters

- Name of the init function – defines the name of an M-file or a MEX file where the initialization code is stored.
- Init function argument – defines the optional arguments that is to be passed to the initialization code.
- Number of analog inputs and outputs – defines the number of inputs and outputs.
- Node numbers - defines the network device as a node.
- Local clock drift – It is the time drift from the nominal time. i.e., actual simulation time.

The local clock drift and offset parameters can be used to introduce jitter and validate the same with the network and controller design.

.

*2) Network block:*

Network block helps in specifying the protocol and can be configured to have packet drops in a specified probability, and if configured packets are dropped randomly with the probability.

- Network type – defines the type of the network and underlying protocol used.
- Network number – it is a network ID and every network should have separate number.
- Number of nodes – defines the number of kernel elements in the network.
- Data rate (bits/sec) – To set network speed.
- Minimum frame (bits) – Messages are constrained to this value.
- Loss probability – probability of packet drop in the network.

Note that the Network block supports various Protocols including Ethernet and CAN networks.

.

*3) Wireless Network:*

This block simulates transfer of a packet in a wireless network with appropriate protocol and depends on the following parameters.

- Network Type – defines the type of network and determines the underlying MAC protocol to be used.
- Network Number – defines the unique id of network.
- Number of nodes – it is a network ID and every network should have separate number.
- Data Rate and Minimum Frame size – defines the bandwidth and the minimum size of each packet.
- Transmit power, signal threshold and path loss exponent helps in signal reception and network availability for the wireless nodes.
- Ack timeout and retry limit helps in retransmission of packets for specific protocols that supports retransmission.
- Optional 'Show Power consumption output port' helps in monitoring the power consumed by the network.

Note that the wireless network supports WLAN, ZigBee and NCM_WIRELESS Networks. 802.11b, 802.15.4 and 802.11n network standards are implemented by the corresponding network types. Wireless Network requires additional two dimensional position input for its operation

*4) Ultrasound Network* This block helps in simulating the ultrasound networks which emit sound in different frequencies, and requires the following parameters.

- Network number and Number of nodes – Unique network ID and number of kernel nodes in the network.
- Ping Length – defines the distance till which the nodes can transfer the packets.
- Speed of sound – Speed of sound in the particular environment (glass/liquid/air).

*5) Other Blocks:* Other blocks include Battery, Receive and send blocks. The battery block utilizes the available power, and is specified the initial operational energy available in the block. When the simulation is run, the available power is consumed and when the all the power is consumed, the corresponding kernel block is switched off. Further operation of the block is aborted, and the system can be simulated for stability and performance when a particular block (Sensor/Controller) is down after all the configured power is utilized. The Receive and Send blocks are now inbuilt in the Kernel block and might not be necessary to run the simulation, however, if custom protocol implementation is required in the Network block to have control over what packet is transmitted, encryption and decryption of packets etc., these blocks prove useful and are necessary.

Each block in the TrueTime tool is useful to simulate different environments, protocols, physical systems and their timing performances, and can prove an inevitable tool for validating timing tolerances.

III. SIMULATION EXAMPLE

Fig. 2 shows the simulation example of a DC-Servo motor [32] using a networked control system using a wireless communication network. It consists of two nodes namely sensor/actuator node and a controller node. The periodical sampling process is done by sensor/actuator node and is sent to controller node where the control task in this node is calculated and the signal is fedback to the sensor/actuator node [37,38]. The simulation is done using WLAN protocol.

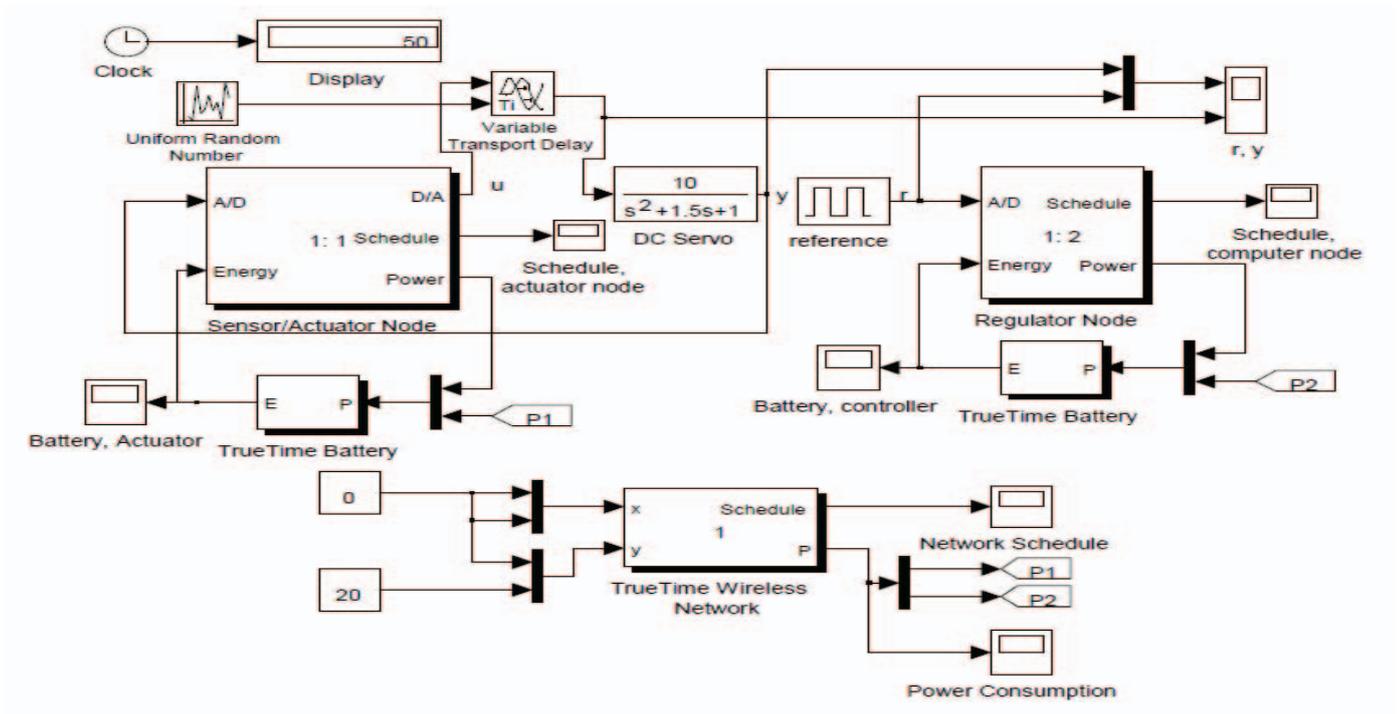

Fig. 2. Block Diagram

The underlying kernel can be initialized with desired scheduling algorithms, so that the system can be validated for different scheduling algorithms, even before practical implementation of scheduling protocols using networks.

TrueTime along with Simulink tool can also validate variable delays introduced as part of network or other modules. The simulation example introduces the variable delay in transmitting the actuator signal to the plant. Network schedule gives the appropriate values for each of the transmitted signals via the network.

Fig. 6 shows the network schedule for the variable delay system. The performance of the system can be obtained from the underlying control algorithm, which again can be specified as an external handler. The timing constraints that are to be specified in the controller algorithm introduces practical delays experienced by the system in real world, and hence is able to validate the timing tolerances of the system.

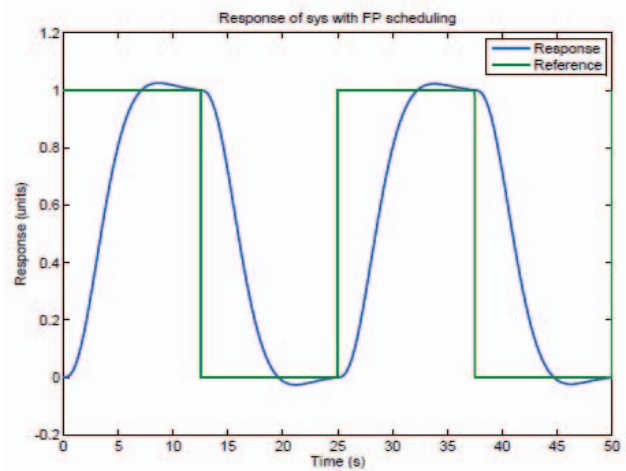

Fig. 3. Response for FP scheduling

In multi-core environments, the number of processors in the system can be set. It also allows us to customize the handlers and interrupts by specifying external handler. TrueTime supports protocols like CAN, Ethernet and other networks and also supports WLAN, Zigbee and NCM_Wireless protocols for wireless networks. These can also be evaluated in a noisy environment by introducing packet drops

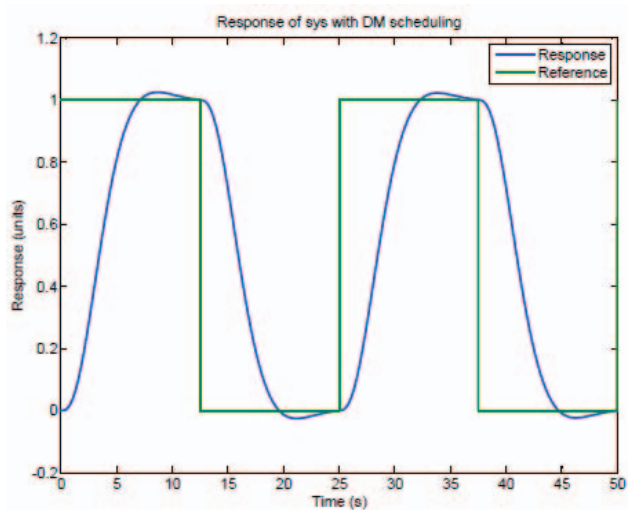

Fig. 4. Response for DM scheduling

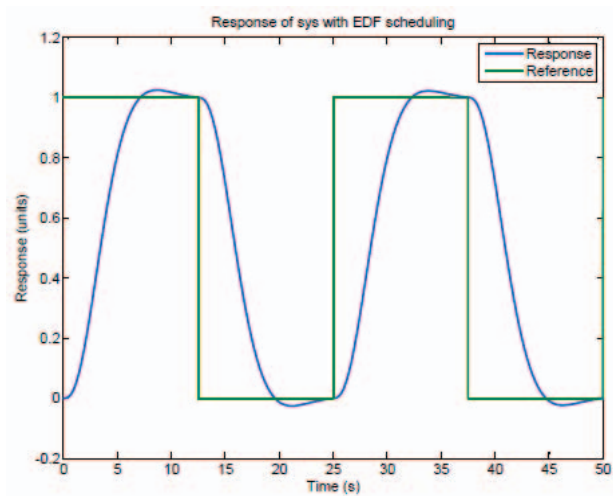

Fig. 5. Response for EDF scheduling

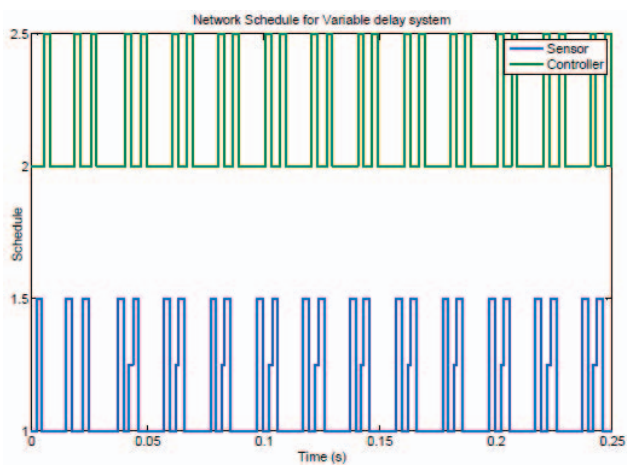

Fig. 6. Network Schedule

## IV. CONCLUSION

TrueTime helps in validating performance of control systems with a given network and helps in simulation of various network protocols. It can be used to study the effect of variation of various network parameters including but not limited to bandwidth, packet loss, and compatibility with different protocols. In combination with Simulink and Matlab toolboxes, it can also simulate real-time systems. TrueTime can be further extended to support more protocols, so that the validation can be done in various environments and the performance of each can be compared. Options to validate various performance measures, timing constraints, various protocols and scheduling algorithms, and options to extend the protocols to implement and include custom protocols, makes it one of the inevitable tools in simulation of CPS.